\documentclass[aps,prb,showpacs,twocolumn,floats,floatfix]{revtex4}
\usepackage{amssymb}
\usepackage{amsbsy}
\usepackage{amsmath}
\usepackage{epsfig}
\usepackage{graphics}
\usepackage{graphicx}

\begin{document}


\title{Approaching the Ground State of a 
Quantum Spin Glass using a Zero-Temperature Quantum Monte Carlo}

\author{Arnab Das} 
\affiliation{The Abdus Salam International Centre for Theoretical Physics,
Strada Costiera 11, 34014 Trieste, Italy \\
\texttt{email: arnabdas@ictp.it}}

\author{Bikas K. Chakrabarti}
\affiliation{Theoretical Condensed Matter Physics Division and Center
for Applied Mathematics and Computational Science, Saha Institute of Nuclear
Physics, 1/AF Bidhannagar, Kolkata 700064, India\\
\texttt{email: bikask.chakrabarti@saha.ac.in}} 

\date{\today}

\begin{abstract}
Here we discuss the annealing behavior of an infinite-range $\pm J$ Ising spin glass
in presence of a transverse field using a zero-temperature quantum Monte Carlo.
Within the simulation scheme,
we demonstrate that quantum annealing not only helps finding the ground state of a classical
spin glass, but can also help simulating the ground state of a quantum spin glass, 
in particularly, when the transverse field is low, much more efficiently. 

\end{abstract}


\maketitle

Quantum annealing (QA) \cite{BC:adbk}-\cite{Somma} is a method of finding the
ground state (minimum energy state) 
of a given classical Hamiltonian employing external 
quantum fluctuations and subsequent adiabatic reduction of them. 
In QA, one is given with a complex classical Hamiltonian
$\mathcal{H}$ whose ground state is to be determined.
In order to introduce the quantum fluctuations 
(necessary for the annealing) into the system, one adds a quantum
kinetic term $\mathcal{H}^{\prime}(t)$ to the classical Hamiltonian,
such that $\mathcal{H}^{\prime}(t)$ and $\mathcal{H}$ do not commute.
Initially, one keeps $\mathcal{H}^{\prime}(t=0) \gg \mathcal{H}$ so that
the total Hamiltonian $\mathcal{H}_{tot}(t) = \mathcal{H}^{\prime}(t) + \mathcal{H}$
is well approximated by the kinetic part only
($\mathcal{H}_{tot}(0) \approx \mathcal{H}^{\prime}(0)$).
The system is initially prepared to be in the 
ground state of $\mathcal{H}^{\prime}(0)$ 
(one chooses $\mathcal{H}^{\prime}(0)$ to have a 
easily realizable ground state).
Now since $\mathcal{H}_{tot}(0) \approx \mathcal{H}^{\prime}(0)$, the 
overlap $|\langle \psi (t)|\phi_{1}(t)\rangle|$ between the lowest eigen-value state
(we will call it instantaneous ground state) $|\phi_{1}(t)\rangle$ 
of the total Hamiltonian $\mathcal{H}_{tot}(t)$ and the instantaneous state
$|\psi(t)\rangle$ of the evolving system will be close to unity at $t = 0$
(since $|\psi (0)\rangle$ is taken to be the ground state of 
$\mathcal{H}^{\prime}(0)$).
If one subsequently reduces $\mathcal{H}^{\prime}(t)$ slowly enough, 
then according to
the adiabatic theorem of quantum mechanics, the overlap 
$|\langle \psi (t)|\phi_{1}(t)\rangle|$ will always stay close to 
its initial value (i.e., unity). Hence
at the end of such an evolution, when $\mathcal{H}^{\prime}(t)$ is reduced to
zero at $t = \tau$, the system will be found in a state $|\psi(\tau)\rangle$ 
with 
$|\langle \psi (\tau)|\phi_{1}(\tau)\rangle| \approx 1$, 
where $|\phi_{1}(\tau)\rangle$
is the ground state of $\mathcal{H}_{tot}(\tau)$, 
which is nothing but the surviving
classical part $\mathcal{H}$ of the Hamiltonian. Thus 
at the end of an adiabatic annealing the system is found in the ground
state of the classical Hamiltonian with a high probability.
Based on this principle, algorithms can be framed to anneal 
complex physical systems like spin glasses
as well as the objective functions of hard combinatorial
optimization problems (like the traveling salesman problem or TSP) 
mapped to glass like Hamiltonians, towards
their ground (optimal) states. 

Here we study the annealing behavior of an 
infinite-range $\pm J$ Ising spin glass
in a transverse field, using a zero-temperature quantum Monte Carlo.
The model and the basic QA scheme for it are introduced in Sec \ref{Model}. 
In Sec. \ref{QMC} we discuss at length the zero-temperature quantum Monte Carlo method 
used here. We discuss the results of QA employed to reach the ground state of the classical spin 
glass in Sec. \ref{Result}. 
We demonstrate in Sec. \ref{QMC-QSG}, how QA can be utilized
to simulate the ground state of a quantum spin glass. 
We conclude with a short summary.

\begin{figure*}
\begin{center}
\resizebox{10cm}{!}{\rotatebox{270}{\includegraphics{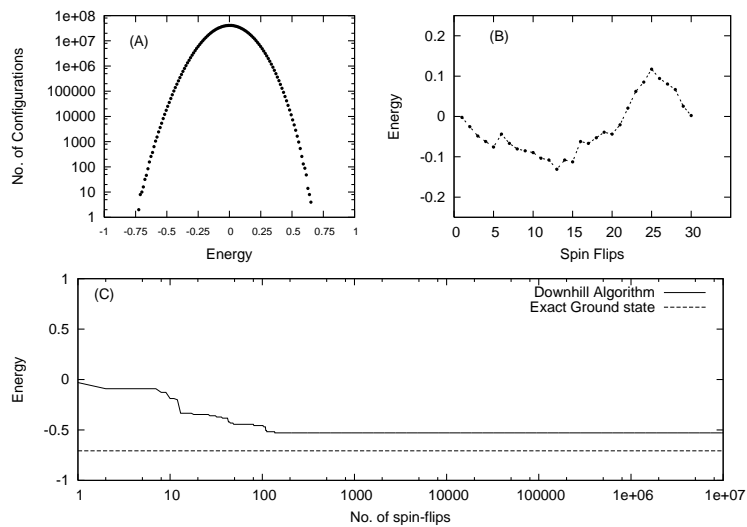}}}
\caption{\small{Here the complexity of finding the ground state for
a typical $30$ spin sample of the $\pm J$ spin glass (Eq. \ref{pmJ}) is
illustrated. In (A) we plot the energy versus number of
configurations for the sample. In (B) we plot the change in energy of the
system as one flips the spins one-by-one, starting from a ferromagnetic 
(all up) state till the mirror symmetric one is reached.
In Panel (C) we show how energy changes with time (spin flips) for
a local downhill optimizer, who flips a spin randomly and accepts the 
move if it reduces the energy. 
}}
\label{pel}
\end{center}
\end{figure*}

\section{Model}
\label{Model} 
We consider an infinite range 
Ising spin system whose Hamiltonian is given by
\begin{equation}
{\mathcal H} = -\sum_{i,j(>i)}^{N} 
J_{ij}\sigma_{i}^{z}\sigma_{j}^{z},
\label{pmJ}
\end{equation}
\noindent where $\sigma_{i}^{z}$ is the $z-$component 
of Pauli spin, representing a classical Ising spin at site $i$ 
and $J_{ij}$'s are random 
variables taking up values either $+1$ or $-1$ with equal probabilities. 
The above Hamiltonian describes a cluster of $N$ Ising spins, each connected 
to all others through exchange interactions of equal strength 
($J = 1$) but random signs. To make the energy extensive in system-size, one
has to scales the energy with a factor of $N^{3/2}$, as done in the rest of the
article. The system is heavily frustrated 
(i.e., no spin configuration can satisfy all the bonds), due to
the presence of both ferromagnetic and antiferromagnetic bonds in random
fashion. The high degree of connectivity 
(i.e., the infinite range of the interactions) adds to the 
complexity of the problem. For such a system, finding the ground state 
spin configuration for any
arbitrary given realization of interactions (the set of $J_{ij}$'s), 
is known to be an NP-hard problem\index{NP-hard problem} \cite{Barahona}. 
In thermodynamic limit the system becomes a non-ergodic spin 
glass below some spin glass temperature $T_G$. To illustrate the complexity
and the difficulty in finding the ground state we consider a typical sample 
with $N = 30$ (we studied many samples, and the features are qualitatively 
the same as we for this one) as shown in Fig. \ref{pel}. The plot of
the number of spin configurations against the energy 
(panel A of Fig. \ref{pel}) of the system clearly
shows that there is a huge entropy barrier between the ground state and
the states which has higher energies. A levels with energy
close to zero has a degeneracy of the order of $10^7$, while the ground state
degeneracy is of the order of unity. This problem could be easily 
overcome and ground state could be reached if the potential energy 
landscape (PEL) had have a more or less monotonic
gradient towards the ground state, as observed in ferromagnetic samples 
(or may be one having few local non-monotonicity for samples with 
a few disorder and short-range interactions). But Here, as 
illustrated in panel (B), the PEL is quite random. 
In the Fig. {\ref{pel}}(B), we plot the energy against spin flip, as we
start from a completely ordered (all up) configuration and flip one-by-one all
the $30$ spins until the mirror-symmetric state (all down). 
In a ferromagnet, a similar plot would generate a convex
curve (with the peak at the zero energy) symmetric about the energy 
axis, reaching the ground state at its both
end. But here we find a completely random profile 
far above the real ground state energy. There is no helpful gradient to 
guide towards the ground state (if one incorporates moves with more than
one spin-flips, then the profile becomes even more rugged). These two 
features, that is, the extremely low entropy of the ground state and the
absence of a consistent gradient towards the ground state in the PEL makes
it absolutely difficult for a random walker to find a ground state
with the help of a naive local minimization algorithm, like the principle of
going down hill, as shown in Fig. {\ref{pel}} panel (C). Here we show the
energy of the system evolved using the random single spin-flip move, and
allowing it to accept the move only if it minimizes the energy. We find
even after $10^{7}$ moves, it fails to reach the ground state. One can
guess that after getting down in energy in first $\sim 100$ steps, it 
gets lost among the huge number of configuration with higher energies 
(which it does not accept) and cannot find a way to reach the sparsely 
occurring lower energy states.

The eigenstates of
${\mathcal H}$ (the basis states) are the direct-products of the
eigenstates of
$\sigma_{i}^{z}$'s. Each basis state represents a distinct spin configuration
of the system.
To perform
zero-temperature 
quantum annealing\index{quantum annealing!of a $\pm J$ Ising spin glass} 
of this $\pm J$ Ising system, 
we add
a transverse field term
${\mathcal H}^{\prime} = \Omega(t)\sum_{i=1}^{N}\sigma^{x}_{i}$ where 
$\sigma^{x}_{i}$'s are $x$-components of Pauli spins which introduces
probability of tunneling between the basis states (classical configurations), 
and $\Omega(t)$ is the strength of the transverse field. The total
Hamiltonian is thus given by

\begin{equation}
{\mathcal H}_{tot} = {\mathcal H} + {\mathcal H}^{\prime}(t)
= -\sum_{i,j(>i)}^{N} 
J_{ij}\sigma_{i}^{z}\sigma_{j}^{z} - 
\Omega(t)\sum_{i=1}^{N}\sigma^{x}_{i}.
\label{Ham-PMJ} 
\end{equation}
  
We start with a high enough value of $\Omega$ initially (at $t = 0$) and
sample the ground state of ${\mathcal H}_{tot}$ using a zero-temperature
quantum Monte Carlo algorithm (discussed below). During sampling, we reduce the 
strength $\Omega(t)$ of the transverse field following a linear
annealing schedule

\begin{equation}
\Omega(t) = \Omega_{0}(1-t/\tau),
\label{Gamma}
\end{equation}
\noindent
where $t$ denotes evolution time. 
At the end of the simulation $(t = \tau)$ 
we are left with the classical
Hamiltonian ${\mathcal H}$ and if $\tau $ is large
enough, the simulated system is finally
found to be in one of its ground state configurations.  
For low values of $\tau$, one generally ends up with
a higher energy configuration.

\section{The Zero Temperature Quantum 
Monte Carlo Method}
\label{QMC}

To simulate the ground state of ${\mathcal H}_{tot}$, we use a zero-temperature
quantum Monte Carlo technique\index{quantum Monte Carlo!zero-temperature}
 \cite {A:Oliveira}. We describe the method to some details here, since
it is not broadly known, and has been implemented so far only to
simulate pure systems with short-ranged interactions. Here we generalize 
the implementation for an infinite-range system with disorders.

In this method one makes a linear
transformation of the form 
\begin{equation}
{\mathcal W} = C{\mathcal I} - {\mathcal H}_{tot}, 
\label{LT}
\end{equation}
\noindent where $C$ is a suitable real constant and ${\mathcal I}$ is the
identity operator, such that the matrix representation of ${\mathcal W}$ 
in the
eigen-basis of ${\mathcal H}$ is non-negative and 
irreducible\index{matrix!irreducible} 
(if such a transformation could not be done for an ${\mathcal H}_{tot}$, 
then this method would not be applicable for it). One can then consider 
${\mathcal W}$ to be the 
transfer-matrix\index{transfer-matrix}
 of a uniform chain 
(with Periodic Boundary Condition(PBC)) of classical plackets, 
where each placket is
nothing but a classical cluster of $N$ mutually interacting Ising spins 
represented by  ${\mathcal H}$.
 
Now the key point is that one can simulate the chain of classical plackets using
the elements of its transfer-matrix\index{transfer-matrix}
 ${\mathcal W}$ and in this simulation the
equilibrium average of any observable (say, energy) related to a single placket
is approximately equal to the expectation value of the observable over the dominant
eigenstate of ${\mathcal W}$. The dominant eigenstate of ${\mathcal W}$ in turn,
 is the ground state of ${\mathcal H}_{tot}$ (due to the form of the linear 
transformation between them). Thus we actually simulate the ground state properties
of ${\mathcal H}_{tot}$ by simulating the chain. 
In the next section we establish the 
scheme in details.


\subsection{Simulation of a chain of classical plackets 
using Transfer-matrix}
\begin{figure}
\begin{center}
\resizebox{4.5cm}{!}{\includegraphics{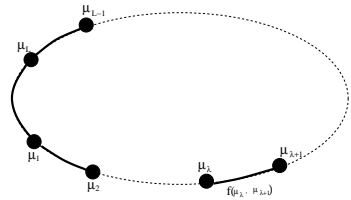}}
\caption{\small{The figure shows the uniform  chain of plackets 
(with periodic boundary condition) used to
simulate the ground state of ${\mathcal H}_{tot}$ (Eq. (\ref{Ham-PMJ})). 
A placket (solid circle) in the chain is basically the cluster of $N$ Ising
spins with a given realization of $J_{ij}$'s, represented by
 ${\mathcal H}$ in Eq. (\ref{Ham-PMJ}). The interactions 
$f(\mu_{\lambda},\mu_{\lambda+1})$ between any two nearest-neighbor-plackets 
are determined by the relation ${\mathcal W}_{\mu_{\lambda}\mu_{\lambda+1}} = 
e^{-\beta f(\mu_{\lambda},\mu_{\lambda+1})}$, where ${\mathcal W}$ is
obtained from ${\mathcal H}_{tot}$ by the linear transformation ((\ref{LT})). 
If the
dimension of each placket is $d$, then the dimension of the resulting chain
is $d+1.$}}
\label{A:plckt-chn}
\end{center}
\end{figure}

In this subsection we demonstrate that the equilibrium averages for a 
single member
of a uniform classical chain (with PBC) is approximately equal to the 
respective
averages (expectation values) over the dominant eigenstate of the 
transfer-matrix\index{transfer-matrix!dominant eigenstate}
of the chain.
Let us consider a uniform chain of $L$ identical classical spin 
clusters (or may be any
 localized discrete degrees of freedom in general) $\mu_i$'s, as 
shown in Fig.~(\ref{A:plckt-chn}). Each of the $\mu_i$'s can be in, say, $p$ 
different states. 
One may note here, that if each placket $\mu$ is a spin
cluster embedded in dimension $d$, then the chain is actually a 
$d+1$-dimensional object. Since the chain is uniform, its Hamiltonian 
will be of the form
$${\mathcal H}_{d+1} = \sum_{\lambda=1}^{L} f(\mu_{\lambda},\mu_{\lambda +1}),$$
\noindent where $f(\mu_{\lambda},\mu_{\lambda +1})$ is a $p\times p$ matrix
whose elements are the possible contributions to the Hamiltonian 
from a pair of neighboring spins, as each of them takes up $p$ different
values independently. The partition function of the chain is thus given by
\begin{eqnarray*}
Z &=& \sum_{\mu_1=1}^{p} ... \sum_{\mu_L=1}^{p}
\exp{\left[-\beta \sum_{\lambda=1}^{L} 
f(\mu_{\lambda},\mu_{\lambda+1})\right]}\\
&=& \sum_{\mu_1=1}^{p} ... \sum_{\mu_L=1}^{p} e^{-\beta f(\mu_{1},\mu_{2})}
\times e^{-\beta f(\mu_{2},\mu_{3})}\times ...\times 
e^{-\beta f(\mu_{L},\mu_{1})}\\
&=& \sum_{\mu_1=1}^{p}...\sum_{\mu_L=1}^{p} {\mathcal W}_{\mu_{1}\mu_{2}}\times
{\mathcal W}_{\mu_{2}\mu_{3}}\times ...\times {\mathcal W}_{\mu_{L}\mu_{1}},   
\end{eqnarray*} 
\noindent where ${\mathcal W}_{\mu_{\lambda}\mu_{\lambda+1}} = 
e^{-\beta f(\mu_{\lambda},\mu_{\lambda+1})}$, $\beta$ being the  
temperature inverse. 
Again, since each of $\mu_{\lambda}$ and $\mu_{\lambda+1}$ can take up
$p$  independent values (i.e., can be in $p$ independent states), 
${\mathcal W}_{\mu_{\lambda}\mu_{\lambda+1}}$ defines
a $p\times p$ matrix ${\mathcal W}$. 
 Hence summing over all the indices from
$\mu_2$ to $\mu_L$ and recalling the rule of matrix multiplication one gets   
$$Z = \sum_{\mu_{1}=1}^{p} ({\mathcal W}^{L})_{\mu_{1}\mu_{1}} = 
Trace{({\mathcal W}^{L})}.$$ The matrix ${\mathcal W}$ is a transfer-matrix
\index{transfer-matrix}
for the chain. If the matrix ${\mathcal W}$ is symmetric then (it is not the
necessary but the sufficient condition) one can write 
$$Z = \sum_{r=1}^{p} (\theta_{r})^{L},$$
\noindent where $\theta_{r}$ are the eigenvalues of ${\mathcal W}$ ordered
by the index $r$, so that $|\theta_{i}| \ge |\theta_{j}|$ if $i<j$. Here a
few points are to be noted. Since all the elements of ${\mathcal W}$ are
strictly positive at any finite $\beta$, the matrix ${\mathcal W}$ is 
both non-negative and primitive \index{primitive matrix} (i.e., there exists
some finite $n$, such that ${\mathcal W}^{n}$ is strictly positive).
Then according to Perron-Frobenius theorem
\index{Perron-Frobenius theorem}
(see \cite{A:Seneta}) the dominant eigenvalue $\theta_{1}$ is strictly positive 
and non-degenerate.  
Thus
\begin{eqnarray*} 
Z &=& (\theta_{1})^{L} + \sum_{r=2}^{p} 
\left(\frac{\theta_{r}}{\theta_{1}}\right)^L \\
&\approx& (\theta_{1})^L
\end{eqnarray*} 
\noindent Here, the leading order error is $(\theta_{2}/\theta_{1})^L$ and
since $\theta_{1}$ is non-degenerate, 
\begin{equation}
 \lim_{L\rightarrow\infty} \left(\frac{\theta_{i}}{\theta_{1}}\right)^L = 0
\label{error}
\end{equation}
\noindent for any $i \ne 1$.

Now, to see how one can simulate the chain using ${\mathcal W}$, one has to note
that the probability that the chain be in a given state $A$ , in which
$\mu_{1} = \mu_{1}(A),\mu_{2} = \mu_{2}(A) ... $ etc, is
$$\qquad
P(A) = \left(e^{-\beta f[\mu_{1}(A),\mu_{2}(A)]}\times ... \times
e^{-\beta f[\mu_{L}(A),\mu_{1}(A)]}\right)/Z $$
\begin{equation}
 = \left({\mathcal W}_{\mu_{1}(A)\mu_{2}(A)}\times ... \times
{\mathcal W}_{\mu_{L}(A)\mu_{1}(A)}\right)/Z 
\label{P-A}
\end{equation}
Thus using the conditions of detailed balance, one obtains 
transition probability from a state $A$ to another state
$B$ given by
\begin{equation}
P(A\rightarrow B) = \frac{{\mathcal W}_{\mu_{1}(B)\mu_{2}(B)}\times ... 
\times{\mathcal W}_{\mu_{L}(B)\mu_{1}(B)}}
{{\mathcal W}_{\mu_{1}(A)\mu_{2}(A)}\times ... \times
{\mathcal W}_{\mu_{L}(A)\mu_{1}(A)}}. 
\label{P-AB}
\end{equation}
\noindent Thus if ${\mathcal W}$ is given, we can simulate the 
equilibrium properties 
(thermal average) of any physical quantity 
 related to a placket $\mu$ in the chain. To obtain 
that, we require to know the probabilities for the placket $\mu$ to be in
its different possible states when the chain is in equilibrium. Let
 $P(\mu = k)$ denotes the probability that the placket is
found in its $k$-th state when the chain is at 
thermal equilibrium
(at a given $\beta$). If the $k$-th state is represented 
by a column vector $|k\rangle$, then
these column vectors satisfy the matrix relation
$$\langle i|{\mathcal W}|j\rangle = {\mathcal W}_{ij},$$
where $\langle i|$ is the transpose of $|i\rangle$ and the
sequence of matrices implies the proper multiplications
between them.

On the other hand, if $|E_1\rangle$ be the dominant 
(normalized) eigenvector of ${\mathcal W}$
corresponding to the dominant eigenvalue $\theta_{1}$, and if
${\mathcal W}$ is hermitian then
one can expand $|E_1\rangle$ linearly in terms of the 
basis vectors as
\begin{equation}
|E_1\rangle = \sum_{k=1}^{p} \gamma_{1}^{k}|k\rangle, 
\label{L-Dcomp}
\end{equation}
\noindent where $\gamma^{1}_{k}$ is the amplitude of the basis state
$|k\rangle$ in $|E_1\rangle$. Thus in the sampling of 
$|E_1\rangle$ using the basis states $|k\rangle$'s, the probability
of occurrence of the state $|k\rangle$ will be $|\gamma^{1}_{k}|^2$.
Now, one can show that 
\begin{equation}
P(\mu = k) = |\gamma^{1}_{k}|^2 + 
{\mathcal O}\left[(\theta_{2}/\theta_{1})^L\right]. 
\label{P-mu-k}
\end{equation}
\noindent The above equation says that one can sample the dominant 
eigenstate $|E_1\rangle$ of
the matrix ${\mathcal W}$ just by sampling its basis states 
(classical configurations of a placket in the chain) according to
the probability of their occurrence in the simulation of the placket
at equilibrium in the chain (using the elements of ${\mathcal W}$ 
itself, as prescribed in (\ref{P-AB})).

To prove equation (\ref{P-mu-k}), 
we take any placket in the chain and call it $\mu_{1}$.
Probability that $\mu_{1}$ is found in the state $|k\rangle$ is
\begin{equation}
P(\mu_{1} = k) = \frac{1}{Z}\left[\sum_{\mu_{2}}\sum_{\mu_{3}} ... 
\sum_{\mu_{L}} 
{\mathcal W}_{\mu_{1}\mu_{2}}{\mathcal W}_{\mu_{2}\mu_{3}} ... 
{\mathcal W}_{\mu_{L}\mu_{1}}\right]_{\mu_{1} = k} $$
$$= \frac{1}{Z}\left({\mathcal W}\right)^{L}_{kk} = 
\frac{\langle k|({\mathcal W})^L|k\rangle}{trace \{({\mathcal W})^L\}}
\label{P-mu-k-Z}
\end{equation}
\noindent Above, we have summed up the probabilities of
all the configurations of the chain, in which $\mu_{1} = k$. 
Now let $|\theta_i\rangle$ ($i = 1,2, ... p$) 
denote the normalized eigenvector of ${\mathcal W}$
corresponding to the eigenvalue $\theta_i$. 
Then one may have a linear transformation\index{linear!transformation} between
$|\theta_i\rangle$'s and $|k\rangle$ of the form
$$|\theta_i\rangle = \sum_{k} \gamma_{k}^{i}|\mu_{k}\rangle$$
\noindent and the reverse transformation
$$|k\rangle = \sum_{i}(\gamma^{\dagger})_{i}^{k}|\theta_{i}\rangle = 
\sum_{i}\gamma^{i\ast}_{k}|\theta_{i}\rangle,$$
\noindent $\gamma$ being an unitary matrix. Hence
\begin{eqnarray*} 
{\mathcal W}^{L}|k\rangle &=& \sum_{i} 
\gamma_{k}^{i\ast}\theta_{i}^{L}|\theta_{i}\rangle\\ 
=> \langle k|{\mathcal W}^{L}|k\rangle &=& \sum_{i}
 |\gamma_{k}^{i}|^2\theta_{i}^{L},
\end{eqnarray*}
\noindent using ortho-normality of $|\theta_{i}\rangle$'s.
\noindent Thus, from equation (\ref{P-mu-k-Z}) we get
\begin{eqnarray*}
P(\mu_{1} = k)\\
&=&\frac{\langle k|{\mathcal W}^{L}|k\rangle}{trace\{{\mathcal W}^{L}\}}\\
&=&\frac{\sum_{i} |\gamma_{k}^{i}|^2\theta_{i}^{L}}{\sum_{i} \theta_{i}^{L}}\\
&=&\frac{\sum_{i} |\gamma_{k}^{i}|^2(\theta_{i}/\theta_{1})^{L}}
{1 + \sum_{i\ne 1} (\theta_{i}/\theta_{1})^{L}} \\
&\approx& |\gamma_{k}^{1}|^2 + {\mathcal O}[(\theta_{2}/\theta_{1})^L],
\end{eqnarray*}  
\noindent which proves equation (\ref{P-mu-k}).  

Thus one can in fact simulate the dominant eigenstate
of any given suitable (hermitian, non-negative and primitive)
$N\times N$ matrix up to a good approximation using the above results.
One has to define a uniform chain (with PBC) of classical plackets, each
having $N$ possible configurations. 
The $i$-th state of a placket corresponds to the $i$-th vector of the basis
in which the given matrix is represented. One then views the given matrix
as the transfer-matrix for a placket in the chain, and simulate the
chain using its elements (as prescribed in (\ref{P-AB})). At equilibrium,
the probability of getting a placket in its $i$-th state is equal to the
modulus square of the weight of the $i$-th basis vector in the representation
of the dominant eigenstate of the given matrix (up to an error of the form
discussed above).

\subsection{Implementation of the Monte Carlo\index{quantum Monte Carlo!zero-temperature}}

We now illustrate the implementation of the above Monte Carlo scheme by
employing it to simulate the ground state of ${\mathcal H}_{tot}$ given in
Eq. (\ref{Ham-PMJ}). Here basis vectors $|k\rangle$'s 
are the eigenvectors of ${\mathcal H}$,
and a classical placket is the cluster of $N$ 
Ising spins with exchange interaction
described by ${\mathcal H}$. Now we make a linear 
transformation\index{linear!transformation} of the form
given in Eq. (\ref{LT}), with $C = N(N-1)/2$. The resulting ${\mathcal W}$ matrix is 
clearly non-negative (since none of its diagonal element are all smaller than 
$N(N-1)/2$ and off-diagonal elements are either $0$ or $\Omega(t)$, 
which we always
take to be positive.). 
Since ${\mathcal H}_{tot}$ connects a basis state to all other
basis states that can be obtained by a single spin flip from it, 
there is no closed
subspace for ${\mathcal H}_{tot}$. Thus ${\mathcal W}$ is also 
irreducible\index{matrix!irreducible}. 
It can be shown that for a non-negative\index{matrix!non-negative} 
irreducible\index{matrix!irreducible} matrix, all the results of
Perron-Frobenius theorem\index{Perron-Frobenius theorem}
 we have used here, holds good \cite{A:Seneta}. 
Besides, ${\mathcal W}$ is of course hermitian. 
Hence we can take ${\mathcal W}$
as a transfer-matrix\index{transfer-matrix}
for the chain. It corresponds to
some interaction 
$f(\mu_{\lambda},\mu_{\lambda+1})$ 
between two neighboring  
($\mu_{\lambda}$ and $\mu_{\lambda+1}$) and
some inverse temperature $\beta$
(not explicitly important here), given by
$${\mathcal W}(\mu_{\lambda},\mu_{\lambda+1}) 
= e^{-\beta f(\mu_{\lambda},\mu_{\lambda+1})}.$$

To simulate the ground state of  ${\mathcal H}_{tot}$ at a given $\Omega$
for a particular realization of $J_{ij}$'s, we construct a uniform
chain of $L$ plackets with PBC. Each placket
is a cluster of $N$ classical Ising spins (described by cooperative
term of ${\mathcal H}_{tot}$) connected through the given
particular realization of $J_{ij}$'s (see (\ref{A:plckt-chn})).  
We start with an arbitrary spin configuration (same for all plackets) and
a given value of $\Omega$. In one Monte Carlo 
step\index{quantum Monte Carlo!zero-temperature} we randomly visit $L$
plackets. At each such visit we make an allowed move 
(a move whose probability is not trivially zero), such that the chain goes
from a state $A$, say, to a new state, say $B$. The probability of acceptance
of the move is nothing but the transition probability 
$P(A \rightarrow B)$ calculated following (\ref{P-AB})
(using the elements of ${\mathcal W}$). While sampling, one can easily
avoid moves whose probabilities are trivially zero (due to the sparsity of the
matrix ${\mathcal W}$) by constructing a more restricted Markov process
to do the sampling \cite{A:Oliveira}.

For doing 
quantum annealing\index{quantum annealing!of a $\pm J$ Ising spin glass}
 of the same system, we start with a high enough
value of $\Omega$ and reduce it very slowly with time $t$ (Monte Carlo step)
following a linear schedule. During visiting different plackets in a given
Monte Carlo step, $\Omega$ is however held fixed. The linear schedule is
specified by $\Omega(t=0) = \Omega_{in}$ and The total number of Monte Carlo
steps executed; $\Omega_{in}$ is linearly reduced to zero with $t$ 
within $95\%$ of the total Monte Carlo steps, we thus have
$\Omega(t) = \Omega_{in}(1-t/\tau)$, where $\tau$ is the annealing time. 

\section{Results and Discussions}
\label{Result}
\begin{figure*}
\begin{center}
\resizebox{9cm}{!}{\includegraphics{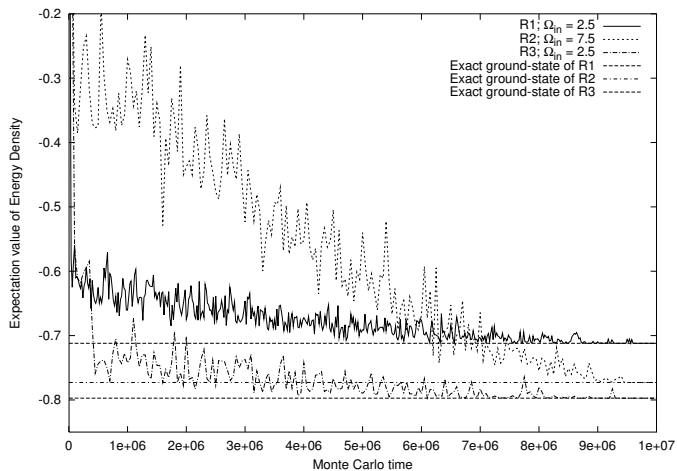}}
\caption{\footnotesize{In the figure, annealing behavior 
of three different randomly
generated realizations (R1, R2 and R3) of $J_{ij}$'s are shown for $N = 30.$ 
In Each case
the system goes to the exact ground state (shown by respective 
horizontal lines) at the end of the annealing. In each case the annealing 
time is $10^7$ Monte Carlo steps, number of plackets  in the chain is $L = 600$,
 and each Monte Carlo step consists of visiting $L$ plackets randomly and 
making a random spin-flip trial there. In each case, the 
transverse field has been reduced to zero from its initial value $\Omega_{in} = 5$ following a 
linear schedule\index{linear!annealing schedule}, within the Monte Carlo steps. 
}}
\label{A:threeR} 
\end{center}
\end{figure*}

We have studied the relaxation behavior of several random $J_{ij}$ 
samples with $N = 30$ for linear annealing 
schedule\index{linear!annealing schedule} 
(we start with an initial transverse field $\Omega_{in}$ and reduce it
linearly with Monte Carlo step, so that it becomes zero before last 
few, $5\%$, steps. 
We observe that for
an annealing of $\sim 10^7$ Monte Carlo steps, the system reaches the true
ground state (determined by an extensive search method) in almost every case,
for a suitably large initial transverse field $\Omega_{in}$. We calculate the
average exchange energy of the chain (over $L$ plackets) in each Monte Carlo
step, and average that over a few $\sim 500$ Monte Carlo steps. The exchange 
energy (as given by ${\mathcal H}$ of Eq. (\ref{Ham-PMJ})
 is not linear in $N$ and 
we have to scale it by a factor $N^{3/2}$ to obtain the 
intensive energy density. In thermodynamic limit, this intensive energy
density approaches the value $-0.7633$\cite{(BC:Parisi)} (our finite size results
shows some fluctuations about that). In Fig. (\ref{A:threeR})   
the relaxation behavior of three typical
 random realizations (R1, R2 and R3) during their 
annealing are shown. We found that 
for doing annealing of a given sample 
within a given number of steps, there is a suitable
range of $\Omega_{in}$. If $\Omega_{in}$ falls below the range, then the
transition probabilities are too low to be able to anneal the system within
the given time. On the other hand, if $\Omega_{in}$ is above the range, then
the rate of change of $\Omega(t)$ is not slow enough to ensure the 
convergence to the ground state finally (i.e., the evolution in no 
more adiabatic). In Fig. (\ref{A:threeR}), the values of respective $\Omega_{in}$'s
belong to the lower end of the respective ranges. The ranges are generally wide
enough, and one can find a $\Omega_{in}$ within the range, just by
a few trials. 
 
\begin{figure*}
\begin{center}
\resizebox{9cm}{!}{\includegraphics{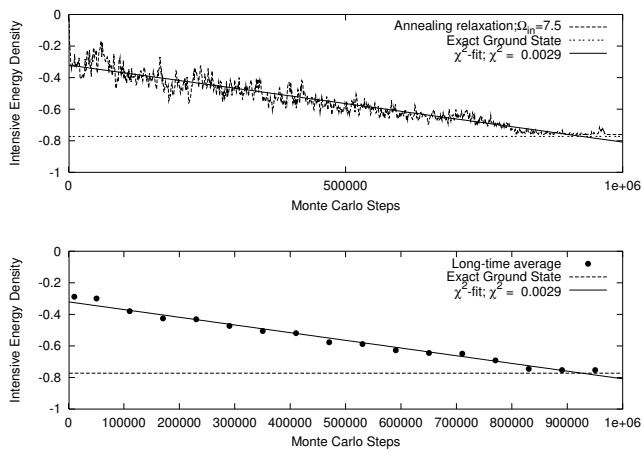}}
\caption{\small{In this figure the 
annealing relaxation ($N = 30$, $L = 600$
and $\Omega_{in} = 7.5$) for a particular 
realization is shown. Here the annealing time $\tau = 10^{6}$
Monte Carlo steps. The upper frame shows the relaxation of intensive energy
density with time, when averaged over small ($\sim 500$) Monte Carlo 
steps. The lower frame shows the same relaxation, 
when the averaging is done over a much
larger number of $(\sim 10^4)$ Monte Carlo steps. 
A linear $\chi^{2}$-fit for the longer time average is shown 
in the lower
part of the figure.}}
\label{A:kifit}
\end{center}
\end{figure*}

The relaxation behavior\index{linear!relaxation} is found to be 
typically ``linear" in the sense that
the long-time averages decrease linearly with time 
(see lower part of Fig. (\ref{A:kifit})). 
The relaxation observed in 
shorter time scale of course shows fluctuations around that 
linear behavior (shown in the upper part of Fig. (\ref{A:kifit})). This linear
nature of relaxation\index{linear!relaxation}
is typically seen independent of the details of the
particular realizations. 

\begin{figure*}
\begin{center}
\includegraphics[width=8.8cm]{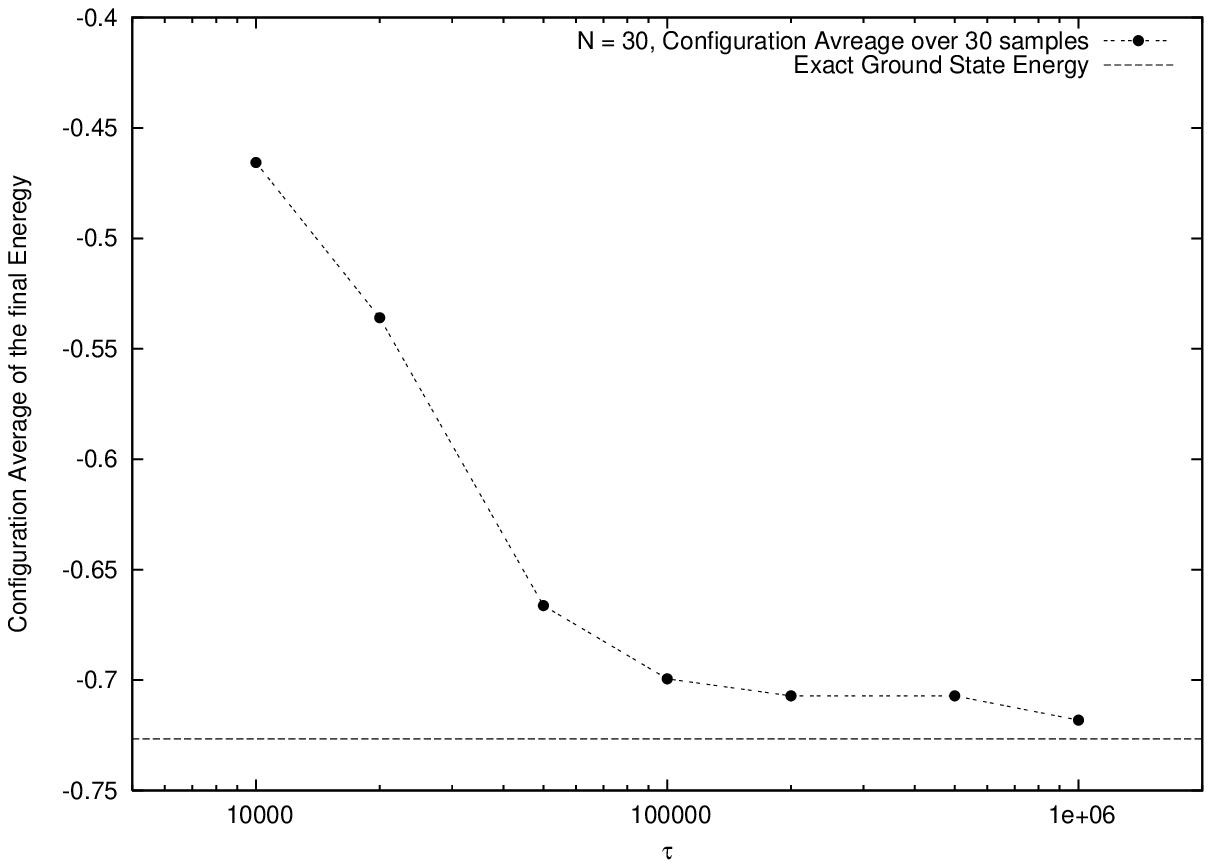}
\end{center}
\caption{\small{Here the variation of the final energy
with the annealing time $\tau$ is shown for the sample 
size $N =30$. Each data point is averaged over the same set of 
$30$ disorder configurations. The transverse 
field $\Omega$ is reduced linearly from $\Omega_{in} = 1.5$
and the replica number is taken to be $L = 600.$ The horizontal
line denotes the exact ground state (averaged over the same set 
of configurations) obtained by an exhaustive search algorithm. 
}}
\label{ZTQMC-Time-Relaxation}
\end{figure*}


\begin{figure*}
\includegraphics[width=7cm,angle=-90]{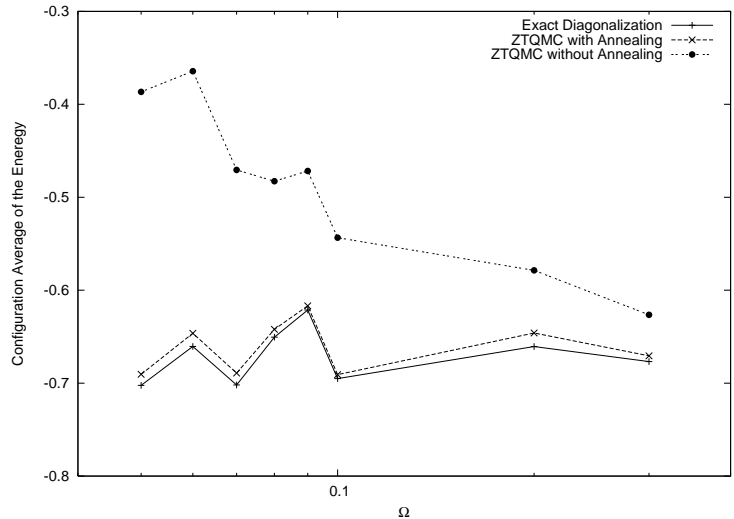}
\caption{\small{In this figure a comparison between the 
results of simulation (with and without pre-annealing) 
 of the ground state of the Hamiltonian (\ref{Ham-PMJ}) 
for different static
values of $\Omega$
and the corresponding exact results obtained 
by numerical exact diagonalization
for the same set of samples. Each data-point represents an average over
the same set of $40$ randomly generated samples of size $N = 20$. The 
total number of Monte Carlo steps is $10^{5}$ for each Monte Carlo 
simulation (including the annealing period for the annealed simulations).
In the figure, he results of simulations with annealing are seen to be much
closer to the exact diagonalization results than those with out annealing.
for lower values of $\Omega$.     
}} 
\label{Simu}
\end{figure*}
   
\section{Better Simulation of Low Kinetic-Energy Quantum States
Using Quantum annealing}
\subsectionmark{Better Simulation Using QA}
\label{QMC-QSG}

In a glassy system, where the potential energy landscape has valleys
separated by huge energy barriers, simulating the ground states 
(and possibly other low-lying states) for low kinetic 
energy (like the ground state for a low value of the
transverse of a transverse Ising spin glass) using a 
zero-temperature quantum Monte Carlo may be very difficult and 
time consuming. This is because, for small kinetic term, the acceptance 
probability may become very
small for higher potential energy states 
energy states, and the system may take a very long time
to get out of a local potential energy minimum in order to
visit other equally relevant lower-potential-energy valleys.
Thus if one gets stuck is a local minimum
far above the ground state, at an early, stage of the simulation,
then it would not be able to reach the low lying valleys, 
whose contributions to the ground state are much more significant.
This can be remedied to some extent 
by annealing the sample quantum mechanically 
starting with a high value of the kinetic energy, and then reducing it slowly
up to the low value at which the simulation is desired. 

For the Hamiltonian given by Eq. (\ref{Ham-PMJ}),
simulation of the ground state 
for a small fixed value of 
the transverse field $\Omega$ 
(using the zero-temperature transfer-matrix
Monte Carlo algorithm described here)
is found to be much closer to the exact
result (obtained using exact diagonalization \cite{Stoer})
when the simulation is done following an annealing 
(reducing $\Omega$ from a high value to the low value
at which the simulation is desired) than that done
directly keeping the value of $\Omega$ fixed to the
low value from the onset. 
We compare the results 
of both kinds of simulations (with and without annealing), 
for several random samples of 
the spin glass for $N = 20$ with
the respective exact 
diagonalization results for them (see Fig. \ref{Simu}).

We conclude summarizing
few points regarding the  
performance of the algorithm described here. 
The algorithm discussed here
is quite general (applicable to any disordered spin system in any
dimension) and may be used for simulating small system-sizes quite
satisfactorily. However, the moves are very restricted, since each
spin flip in a placket requires the two nearest-neighboring plackets
to be in the same configuration with it. For large system sizes, 
this is too restrictive a condition to move freely enough through the
configuration space to procure a satisfactory sampling rate.
In addition, since the acceptance probability
for higher potential energy configurations 
(like most other zero-temperature quantum 
Monte Carlo algorithms) depends on the magnitude of the
kinetic term, it is hard to simulate the ground state
for low values of the kinetic term. 
We have shown how quantum annealing can be utilized
in overcoming this difficulty (at least partially). This remedy is
is expected to work also for other zero-temperature
quantum Monte Carlo methods in principle.

We thank M. J. de Oliveira for providing us with the basic code for
one-dimensional pure Ising chain in transverse field. A. Das is also
grateful to G. Santoro and E. Tosatti for useful discussions.

\end{document}